# Post-Processing Free Spatio-Temporal Optical Random Number Generator Resilient to Hardware Failure and Signal Injection Attacks

Mario Stipčević, [1,2*] and John Bowers,[1]

[1]*Department of Electrical and Computer Engineering, University of California Santa Barbara, Santa Barbara, USA*
[2]*Department of Experimental Physics, Ruđer Bošković Institute, Zagreb, Croatia*
[*]*mario.stipcevic@irb.hr*

**Abstract.** We present a first random number generator (RNG) which simultaneously uses independent spatial and temporal quantum randomness contained in an optical system. Availability of the two independent sources of entropy makes the RNG resilient to hardware failure and signal injection attacks. We show that the deviation from randomness of the generated numbers can be estimated quickly from simple measurements thus eliminating the need for usual time-consuming statistical testing of the output data. As a confirmation it s demonstrated that generated numbers pass NIST Statistical test suite without post-processing.

©2014 Optical Society of America

**OCIS codes:** (270.5585) Quantum information and processing; (270.5570) Quantum detectors; (270.5568) Quantum cryptography.

**References and links**

1. P. Hellekalek, "Good random number generators are (not so) easy to find", Math. Comp. Simul. **46,** 485-505(1998).
2. I. Goldberg, D. Wagner, "Randomness in the Netscape Browser", Dr. Dobb's, January 1996.
3. T. Click, A. Liu, G. Kaminski, "Quality of Random Number Generators Significantly Affects Results of Monte Carlo Simulations for Organic and Biological Systems", J. Comp. Chem. **32**,513-524(2011).
4. A. Proykova, "How to improve a random number generator", Comp. Phys. Comm. **124**,125-131(2000).
5. A. Stefanov, N. Gisin, O. Guinnard, L. Guinnard, H. Zbinden, "Optical quantum random number generator", J. Mod. Opt. **47**, 595-598 (2000).
6. T. Jennewein, U. Achleitner, G. Weihs, H. Weinfurter, A. Zeilinger, "A Fast and Compact Quantum Random Number Generator", Rev. Sci. Instrum. **71**, 1675-1680 (2000).
7. A. Figotin, A. Y. Gordon, S. A. Molchanov, V. P. Popovich, J. E. Quinn, G. N. Stetsenko, N. M. Stravrakas, I. M. Vitebskiy, "A random number generator based on spontaneous alpha-decay", PCT application WO0038037A1.
8. M. A. Wayne, E. R. Jeffrey, G. M. Akselrod and P. G. Kwiat, "Photon arrival time quantum random number generation", J. Mod. Opt. **56**, 516-522(2009)
9. M. Stipčević, B. Medved Rogina, "Quantum random number generator based on photonic emission in semiconductors", Rev. Sci. Instrum. **78**, 045104:1-7 (2007).
10. J.G. Rarity, P.C.M. Owens, P.R. Tapster, "Quantum random-number generator and key sharing", J. Mod. Opt. **41,** 2435-2444 (1994).
11. R. Davies, "Exclusive OR (XOR) and hardware random number generators", February 28, 2002, URL: http://www.robertnz.net/pdf/xor2.pdf
12. J. von Neumann,"Various techniques for use in connection with random digits", von Neumann Collected Works, vol. 5, Pergamon, pp. 768-770 (1963).
13. M. Stipčević, D. J. Gauthier, "Precise Monte Carlo Simulation of Single-Photon Detectors", Oral presentation, Proc. SPIE Defense, Security and Sensing, 29 April - 3 May 2013, Baltimore, Maryland, USA.
14. D. E. Knuth, The art of computer programming, Vol. 2, Third edition, (Addison-Wesley, Reading, 1997).
15. R. Shaltiel, "Recent developments in explicit constructions of extractors". Bull. EATCS, **77,** 6795 (2002).
16. R. Shaltiel "How to get more mileage from randomness extractors", Random Struct. Algorithms, **33**, 157-186(2008).

17. B. Chor, O. Goldreich, J. Hasted, J. Freidmann, S. Rudich, and R. Smolensky, "The bit extraction problem or t-resilient functions" 26th Annual Symposium on Foundations of Computer Science (FOCS), pages 396-407, IEEE 1985.
18. P. Lacharme, "Analysis and Construction of Correctors", IEEE Trans. Inform. Theor., **55**, 4742-4748 (2009).
19. Robert Davies, Statistics Research Associates Limited, 8 Bristol Street, Island Bay, Wellington, 6023, New Zealand (private communication 2013/2014).
20. A. Rukhin, J. Soto, J. Nechvatal, M. Smid, E. Barker, S. Leigh, M. Levenson, M. Vangel, D. Banks, A. Heckert, J. Dray, S. Vo, "A Statistical Test Suite for the Validation of Random Number Generators and Pseudo Random Number Generators for Cryptographic Applications", NIST Special Publication 800-22rev1a (dated April 2010). URL: http://csrc.nist.gov/groups/ST/toolkit/rng/documents/sts-2.1.1.zip

**1. Introduction**

The ability to generate random numbers is an important resource in many areas of science and technology: computer security, cryptography, probabilistic computation (Monte Carlo), over-Turing computing techniques (e.g. randomized algorithms), simulations, labeling of prepaid and gift cards, industrial testing, online hazard games and automata, scientific research etc. The vast majority of today's computers are deterministic (e.g. PCs, tablets and mobile phones), and so they can not *per se* create random numbers: that task is left to a random number generator (RNG). Random number generators are usually categorized as either pseudo random (PRNG) or physical or "true" random number generators (TRNG).

A PRNG is a mathematical formula, or more generally a deterministic algorithm which, starting from a certain initial number (seed) that defines the initial state, produces a string of numbers that looks random in the sense that it possesses a certain set of desirable statistical properties, but in fact is completely deterministic and highly losslessly compressible by definition [1], neither of which should be a characteristic of a truly random sequence.

Although there is no criterion for an algorithm to be named a "PRNG", the only undisputable characteristic of any PRNG is that it is *provably non-random* because it is already known how to predict all the numbers in the pseudo random sequence, namely by using the very PRNG algorithm. Nevertheless, PRNGs are very frequently used: their popularity stems from the fact that they can be realized as a piece of software and run on a computer or programmable devices (mobile phone, smart card, etc.) thus offering an illusion that the device now also has access to random numbers without any cost in additional hardware! However, while algorithmically generated pseudo random numbers can be used for some applications, they are by construction deterministic and therefore, at least in theory, predictable which makes them risky for use in cryptography [2] as well as frequent cause of erroneous results in statistical calculations and simulations [3-4].

A non-determinism is sought in physical RNGs, notably quantum random number generators (QRNG) that extract random numbers by performing repeated measurements on a certain, specially prepared quantum systems [5-9]. The rationale behind QRNGs is that quantum theory allows the existence of fundamentally random and unpredictable physical processes such that one can, in principle, extract truly random numbers from them.

This paper is organized as follows. First, we present an original analysis of two well known optical quantum random number generating principles. Next, based on gained insights, we construct a new bit extracting principle with significantly improved characteristics. Finally, we build and test a physical QRNG based on that principle, using two home-made single-photon detectors described elsewhere [13], and logic circuits programmed within the Altera MAX3000 family CPLD chip. Notable novelty in our approach to random number generation is that we *prove* limits of deviation from randomness by calculating them from a set of simple measurements.

## 2. State tomography spatial method (BSR)

A well known beam splitting principle of generating random binary numbers (bits) [10], [5] is shown in Fig. 1. A non polarized light described by $\psi = \frac{1}{\sqrt{2}}(|H\rangle + |V\rangle)$ is incoming onto a balanced linearly polarizing beam splitter (PBS). A source of such light could, for example, contain many atoms that emit photons independently of each other at random times or a well-saturated laser in which the photon emitting time information is erased. The action of PBS can be described by the operator $\frac{1}{\sqrt{2}}(|H\rangle\langle H| + |V\rangle\langle V|)$ with understanding that the horizontal linear polarized component will exit on one output arm of the PBS and vertical will exit on the other. When non-polarized light enters an ideal PBS followed by two identical detectors D0 and D1, as shown in Fig. 1, there will be an equal probability of a photon being detected by either detector. A photon detection by the detector D0 is defined as generation of bit value "0" whereas a detection by D1 is defined as generation of a bit value "1". Because the value of a random number is identified with the position at which the photon is detected, this method is sometimes referred to as "spatial".

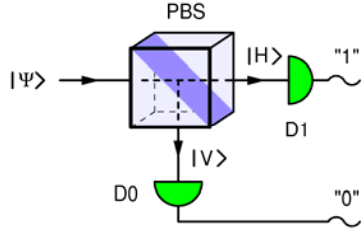

Fig. 1. Beam splitter quantum random number generator. Single-frame excerpts from video recordings of metallic objects concealed by opaque plastic tape. (a) Utility blade. (b)

The setup shown in Fig. 1 generates random outcomes due to the fact that the quantum mechanical system being measured contains no information that could influence which axis the photon state is (more likely) to project. Randomness is therefore a consequence of no information being transferred between the system (setup) and the photon. In fact quite generally, if one imagines a RNG as a "black box", bits coming out of that box must not contain any information whatsoever about what is inside the box simply because being random they do not contain any information about anything.

Even though the described generating method in theory generates perfect random numbers, we will show that it is very sensitive to small imperfections in the experimental setup. Bit sequences generated by any physical RNG generally feature two types of deviations from randomness which seem unavoidable: bias and correlations among bits. Bias is a measure of uneven probability of generating ones and zeros and is defined as:

$$b = \frac{p(1) - p(0)}{2} = p(1) - \frac{1}{2}. \qquad (1)$$

Following the above definition, an estimate of bias from data is obtained as:

$$\hat{b} = \frac{\sum_{i=1}^{N} x_i}{N}. \qquad (2)$$

with the variance of $1/(2\sqrt{N})$. Sources of bias in the BSR method are: (1) unbalanced beam splitter; and (2) uneven detection efficiencies of the two detectors. In the BSR random number generator, it is virtually impossible to eliminate bias by shear manufacturing precision of the components to a level that would not be (easily) detectable by statistical tests. Bias itself can be strongly reduced or eliminated altogether by simple postprocessing techniques [11].

However, exact de-biasing algorithms (such as Von-Neumann [12]) work *only* if there are no correlations among bits or if they are negligible.

Unlike bias, which can be expressed by a single number, correlations among bits can be arbitrarily complex and generally cannot be expressed by a single quantity. Nevertheless the description can be simplified by realizing that memory effects in detectors, that cause correlations among bits, are strongly localized in time [13]. If the mean period between detections is long enough, only neighboring bits are non-negligibly correlated. Under that assumption, bit generation is a Markov process and correlation among bits can be fully characterized by only the serial autocorrelation coefficient with lag 1. Serial autocorrelation coefficients are defined as (see Ref. [14]):

$$a_k = \frac{\sum_{i=1}^{N-k}(x_i - \bar{x})(x_{i+k} - \bar{x})}{\sum_{i=1}^{N-k}(x_i - \bar{x})^2} \quad (3)$$

where $a_k$ is a serial autocorrelation coefficient with lag $k$ and $N$ is number of bits in the sequence. For a Markov process $a_k = a_1^k$. For finite $N$, an estimate given by Eq. (3) has 1 sigma Gaussian statistical error of $1/\sqrt{N-k}$.

In the BSR method, correlations among bits are caused mainly by dead time and afterpulsing in detectors. To see that, let us a suppose that faint continuous Poisson random light, such as for example generated by an LED or well saturated laser, is shone upon the beamsplitter causing photons to be detected by each detector with a mean detection period $\tau$. Let us further suppose that the two detectors have identical dead time $\tau_d$. In case that the next photon "arrives" in less than $\tau_d$ after the previous one, it can either hit the detector that is in the dead state or would get detected by the other detector thus contributing to the negative autocorrelation of the sequence of generated bits. The leading (in magnitude) is autocorrelation coefficient $a_1$, henceforth denoted $a$ and referred to as "autocorrelation". Its magnitude is approximately equal to probability $p_d$ that a photon "falls" into the dead time:

$$a = -p_d = e^{-\tau_d/\tau} - 1 \approx -\frac{\tau_d}{\tau} \quad (4)$$

where the last approximation is valid for $\tau_d \ll \tau$. In a realistic case, the mean photon frequency could be 1 MHz (i.e. $\tau$ = 1000 ns) and dead time $\tau_d$ = 40 ns leading to $a \approx$ -0.04: an imperfection statistically detectable with only a few hundred generated bits! Even with an unrealistically short dead time, at a sufficiently high detection (i.e. bit production) rate, the correlations become un-tolerably high. The highest counting rate of the detector is $1/\tau_d$ and in that limit $a_1$ is approaches -1. To achieve lower correlation one needs to operate at a lower detection frequency, but then only a small portion of detector capabilities can be exploited for random number generation. In our previous work [13] we showed that correlation does not vanish in the low detection frequency end either: instead, it asymptotically reaches the value of the afterpulsing probability $p_a$ which is usually on the order of $0.01 - 0.1$.

In previous work [13] we have already noted that positive and negative correlation mechanisms cancel each other at a certain detection rate. Since the two processes are independent of each other, the resulting correlation $a$ is a sum of the two contributions:

$$a = p_a - \frac{\tau_d}{\tau}. \quad (5)$$

and vanishes for a detection rate $f_0 = 1/\tau = p_a/\tau_d$. For the detectors used in [13], identical to the ones used in this study, $p_a = 0.031$ and $\tau_d$ = 24 ns yielding $f_0 \approx$ 1.3 Mcps, which is quite low with respect to their counting capability of over 25 Mcps.

In this work we go further by noting an effect that, in principle, enable total elimination of the autocorrelation. First we note that by ignoring detections that came from one detector in less than a prescribed blanking time $\Delta t$ after a detection by the other detector, helps to reduce correlations significantly. Specifically, in order to completely remove the anti-correlation caused by the dead time, it would be enough to take $\Delta t$ any higher than the dead time because then the dead time would have no effect on photons selected for generation of random numbers. However, positive autocorrelation due to afterpulses appearing after $\Delta t$ would be left over causing a small positive autocorrelation. In order to overcame that, we note that by taking a suitable $\Delta t$, slightly smaller that the dead time, one can obtain, in principle, an arbitrarily high detection frequency $f_0$ at which the two competing effects (negative and positive) cancel each other. Note that below $f_0$ the total autocorrelation is positive due to dominant effect of the afterpulsing, whereas above $f_0$ autocorrelation turns negative due to dominant effect of the dead time. The circuit shown in Fig 2. provides the function of the BSR method with blanking (omitting) of any event(s) that come sooner than $\Delta t$ after the previous one. Due to the discrete delay times available in the CPLD chip, the blanking period $\Delta t$ is adjustable in the range from 5.6 ns in steps of 4.0 ns, while dead times of the two detectors D0 and D1 used in this work are fixed to 24 ns.

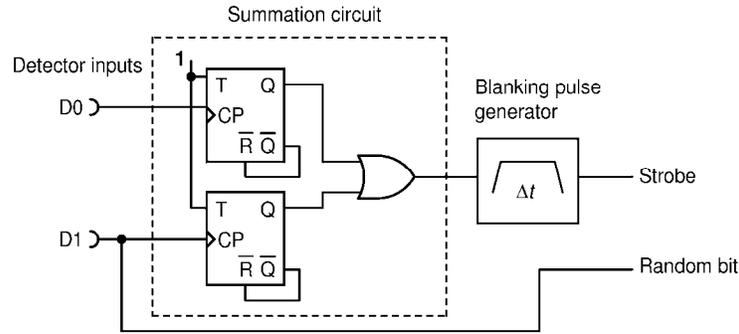

Fig. 2. Logic circuit that implements BSR method. Each flip-flops generates a narrow pulse (~ 2 ns) following a (positive going edge of a) pulse from the respective detector. The OR gate interleaves in time pulses from the two flip-flops and uses that to trigger a blanking pulse of duration $\Delta t$. width  sequences  and The Strobe output generates one short pulse whenever a pulse from a detector arrives except if two consecutive are closer then $\Delta t \approx 2 p_w$ in time in which case only first pulse causes Strobe signal and the other is suppressed.

By checking the higher-lag autocorrelation coefficients we find that a detection rate of 10 Mcps is a good trade-off between a high bit production rate and approximate Markov chain characteristic of the generated sequence. The lowest autocorrelation at 10 Mcps is obtained for $\Delta t = 17.6$ ns and is equal to $(280 \pm 32) \cdot 10^{-6}$. Using that value of the blanking period, the autocorrelation as a function of detection rate has been evaluated according to Eq. 3, with statistics of $N = 10^9$ bits. The result shown in Fig. 3. confirms the expected sign change near 10 Mcps.

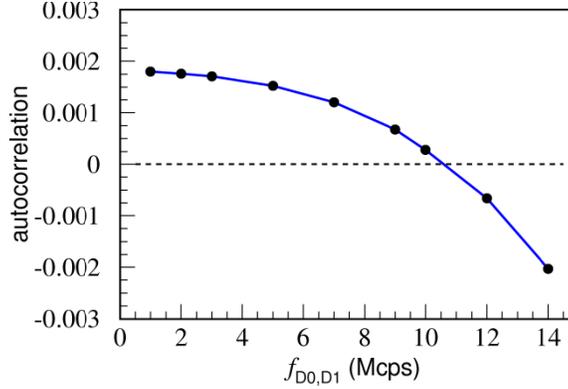

Fig. 3. Autocorrelation coefficient for the BSR method a as a function of the detection frequency $f$ of detectors D0 and D1, for the blanking time $\Delta t = 17.6$ ns.

## 3. Photon pair waiting times difference method (T1T2)

The biggest challenge in realization of a good quantum (or any other physical) RNG is that it is difficult to realize a setup close to the theoretical idealization, especially if there is anything that must be adjusted prior to use. For example to adjust bias in BSR methods to within 1/$N$ one must generate at least ~$N^2$ bits to test the adjustment and is therefore faced with an insurmountable time consuming task. Even if sufficiently fine adjustment could be done, there is a question of whether it would stay stable over time withstanding temperature, power supply and other variations as well as aging and wear of components.

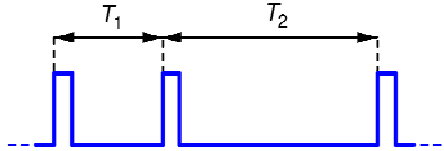

Fig. 4. Illustration of the T1T2 method. Three subsequent random events (detected photons) define two intervals: $T_1$ and $T_2$. If $T_1 > T_2$ then 0 is generated, if $T_2 > T_1$ then 1 is generated whereas if $T_1 = T_2$ then events are skipped and no random bit is generated.

Therefore, especially valuable are random number generating methods that do not require any adjustments. One such method has been proposed in [9]. A time-wise random stream of electrical (logic) pulses is obtained from a random event generator (REG) which generates electrical pulses whose waiting-times obey an exponential probability density function (p.d.f). Three consecutive pulses from the REG, as shown in Fig. 4, are used to define time intervals $T_1$ (between the first two) and $T_2$ (between second and third). A random bit is then generated by comparing the two intervals: if $T_1 > T_2$ then value 0 is generated; if $T_2 > T_1$ then value 1 is generated; if $T_1 = T_2$ (within the measurement precision) then no bit is generated. In order to maximize the bit efficiency, the third event is taken as the first event of the next triplet, thus two events are spent to generate one random bit. If the REG source is stationary and memoryless, which is indicated by an exponential time interval distribution, the bits will be uncorrelated and, due to exchangeability of definitions of 0 and 1, the bias will be zero.

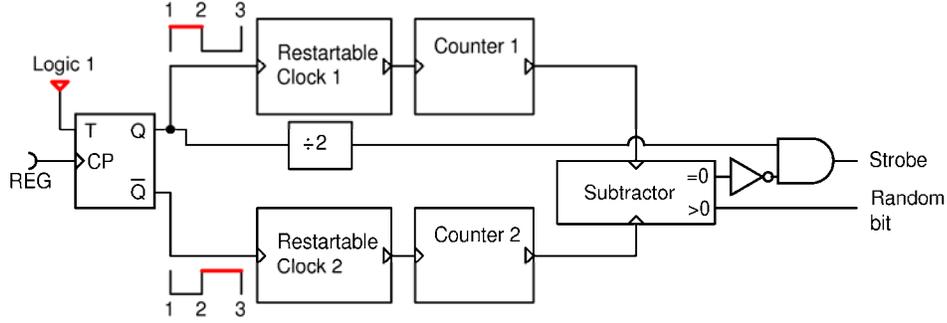

Fig. 5. The T1T2 principle QRBG. Time-wise random pulse train from REG toggles the T-type flip-flop such that the period between the first and second event is measured by the Restartable Clock 1 and the period between second and third event is measured by the Restartable Clock 2. Accumulated counts in the counters are subtracted and a random bit is generated only if the two periods are not equal.

The crucial insight achieved in [9] is that the clock which measures photon time intervals must be started in synchronization with beginning of each interval, otherwise the method would produce correlated bits even if fed by perfectly random events. This was not understood in previous art, like for example in [7] where the clock was free-running which must have yielded correlated bits, but it was not noted probably because clock frequency (~10MHz) was much higher than the source mean frequency (~10kHz) and in that case correlations are small. Correlations rise quickly as the ratio of frequencies of clock and source of random events becomes smaller [7]. The improved T1T2 circuit, shown in Fig. 5, is theoretically exact regardless of the ratio of frequencies of the time measuring clock and REG. Afterpulsing in detectors causes some correlation, but much smaller than in BSR method because they appear at random times and merely slightly modify the exponential distribution.

## 4. Combined spatio temporal method with improved randomness (COMBO)

The two above described random number generating methods offer a range of randomness quality, bits-per-photon efficiency and resilience to hardware imperfections. However, at the present level of technology, even with the most optimal method, the leftover randomness imperfections are typically larger than what is acceptable by general applications. As illustrated there is always some hardware detail that limits the randomness and which cannot be further improved with a given level of hardware technology.

In order to arrive to a better randomness of generated bits, one can use deterministic postprocessing (extractor algorithms [15-16], resilient functions [17-18]) or non-deterministic postprocessing (i.e. using additional sources of entropy). Postprocessing comes with a price of additional hardware and/or software resources. More importantly, complex postprocessing may render it impossible to find a mathematical relation between parameters of the hardware and deviation from randomness, which is a basis for scientific provability of randomness of the generator. In order to simplify postprocessing or avoid it altogether, it is therefore legitimate to ask whether a better bit extraction method can be construed that would both feature lower sensitivity to hardware imperfections and provability of randomness of the extracted bits.

To that end we proceed by noting that in the BSR method (explained in the previous section), there are two independent random processes, each of which allows for random bit extraction. Namely, the time *when* a photon is detected (by either detector) is, at least in theory, completely independent of *where* it is detected (by which detector). Therefore, one can exploit *temporal* information of a train of photons detected by either detector D0 or D1 to generate a sequence of random bits via T1T2 method, and at the same time use *spatial* information of the same detected photons to generate an independent sequence of random bits via BSR method. This method we name "COMBO" as it combines the two random number generating principles.

The optical part of the COMBO RNG is shown in Fig. 6. It contains two near-identical photon counting detectors D0 and D1. It is basically the BSR setup (Fig. 1) with adjustable ratio of zeros to ones.

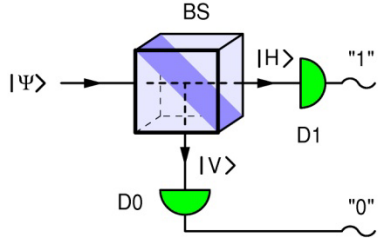

Fig. 6. Optical part of the COMBO quantum random number consists of the beamsplitter PBS, detectors D0 and D1, fixed neutral filter NDF and variable neutral filter VNDF for bias adjustment.

For the random number generation, we used variable-power CW light source made of a red LED (Hamamatsu L7868, $\lambda = 670$ nm, $\Delta\lambda = 30$ nm FWHM) supplied by an adjustable current source and coupled to a single mode fiber (custom shielded SM600 fiber from Thorlabs). The beam splitter BS is a fusion fiber 50:50 splitter (Thorlabs FC632-50B-FC). The neutral density filter NDF reduces intensity of light by a factor of 0.9 whereas the variable filter VNDF making possible a fine adjustment of the bias $b_S$ of the BSR section down to $(0.5 \pm 0.001)$ with stability better than $\pm 0.0005$ during experiments.

Photon detections from the two detectors are combined by the electrical circuit shown in Fig. 7, which functions as follows. In order to extract the timing information, pulse trains from the two detectors are interleaved by a summing circuit to form a single train of pulses which is then processed according to the T1T2 method with the sub-circuit displayed in Fig. 5. This yields one string of random bits, let us denote it with ***T*** for "time". At the same time, the two pulse trains are processed according to the BSR method by the sub-circuit shown in Fig. 2. This yields the other string of random bits related to the space information, let us denote it with ***S*** for "space". Each of these two independent bit strings will have its own bias and autocorrelation, let us denote them as $(b_T, a_T)$ and $(b_S, a_S)$ respectively.

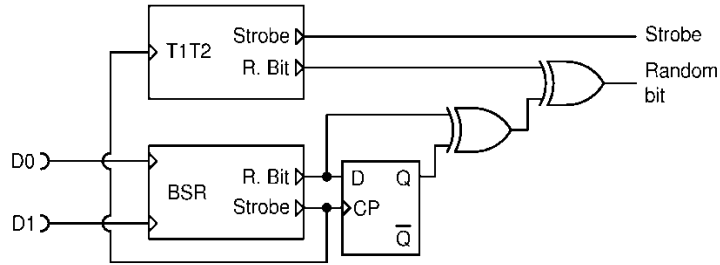

Fig. 7. Functional schematic of the circuit for realizing the COMBO random number generating method.

Since, in principle, the BSR method produces one bit per detected photon and T1T2 produces one bit per two detected photons, we need to decide how to combine these two bit streams. One way is to omit every second BSR bit: this would result in a string with much lower (squared) auto correlation but would not affect the bias. In order to improve both on bias and correlation, the COMBO method uses XOR of non-overlapping pairs of consecutive bits from the BSR method yielding new string $Y(b_Y, a_Y)$. This is accomplished by the D-type flip-flop and XOR gate XOR1 shown in the Fig. 7. Finally the two random strings $T$ and $Y$ are XOR-ed (by the gate XOR2) to yield the final string $C(b_C, a_C)$. In order to calculate estimates of $b_C$ and $a_C$ we use two assumptions discussed above: (1) both BSR and T1T2 methods are Markov processes; (2) random strings generated by two methods are independent of each other. Both assumptions are checked by the data themselves. Note that the T1T2 section is only fed by events passed through the blanking filter of the BSR, even though blanking is not necessary for that section, in order to prevent too often clocking of the BSR section that would induce positive correlation in the string $S$.

For generation of random numbers, the power of the CW source is adjusted such that each detector counts with an average frequency of (10.00±0.05) Mcps, in total 20 million random events per second. In principle this would result in random bit generation rate of 10 Mbps since the bit production is clocked by T1T2 section which uses 2 detections per bit, however due to the blanking the bit rate is approximately 8.0 Mbit/sec.

In order to estimate quality of randomness at the output (sequence $C$) sequences $S$ and $T$ were initially collected from the circuit shown in Fig. 7. Statistics of $2 \cdot 10^9$ bits for string $S$ and $1 \cdot 10^9$ for string $T$ were collected. Statistical analysis according to Eqs. (2) and (3) yielded the following estimates: $\hat{b}_S = (227 \pm 16) \cdot 10^{-6}$, $\hat{a}_S = (-149 \pm 32) \cdot 10^{-6}$ and $\hat{b}_T = (-125 \pm 16) \cdot 10^{-6}$, $\hat{a}_T = (48 \pm 32) \cdot 10^{-6}$. We find that second and further autocorrelation coefficients ($k \geq 2$) for either of the two strings are consistent with the Markov hypothesis within statistical uncertainty. For the string $Y$, which is derived from string $S$, the following relation holds [11], [19]:

$$b_Y = -2b_S^2 - 2a_S\left(\frac{1}{4} - b_S^2\right) \approx -2b_S^2 - \frac{a_S}{2} \tag{6}$$

$$a_Y = 2\frac{a_S(1-a_S)b_S^2}{1 - 2(1-a_S)\left(\frac{1}{4} - b_S^2\right)} \approx 4a_S b_S^2 \tag{7}$$

where the approximations are valid when $a_S \ll 1$ and $b_S^2 \ll 0.25$ which is satisfied in our case. This yields: $\hat{b}_Y = (75 \pm 16) \cdot 10^{-6}$, $\hat{a}_Y = (0 \pm 21) \cdot 10^{-6}$. Again, higher lag correlations are consistent with zero within statistical error margins.

Bit-by-bit XOR-ing of two independent Markov sequences $T$ and $Y$ yields the final sequence $C$ with bias and correlation given by [11]:

$$b_C = 2b_T b_Y \tag{8}$$

$$a_C = a_T a_Y + 4(a_T b_Y^2 + a_Y b_T^2) \tag{9}$$

The above relations may give an over-optimistic result if the two sequences are correlated, that is if normalized cross-correlation coefficients defined as:

$$a_{TY}(k) = \frac{\sum_{i=1}^{N-k}(t_i - \bar{t})(y_{i+k} - \bar{y})}{\sqrt{\left(\sum_{i=1}^{N-k}(t_i - \bar{t})^2\right)\left(\sum_{i=1}^{N-k}(y_i - \bar{y})^2\right)}} \tag{10}$$

are non-zero. Estimates of cross-correlation coefficients for lag $k$ in the range $[-6, 6]$ for a $3 \cdot 10^9$ bits long strings, shown in the Fig. 8. with 1 sigma Gaussian error bars are consistent with zero thus confirming the non-correlation hypothesis.

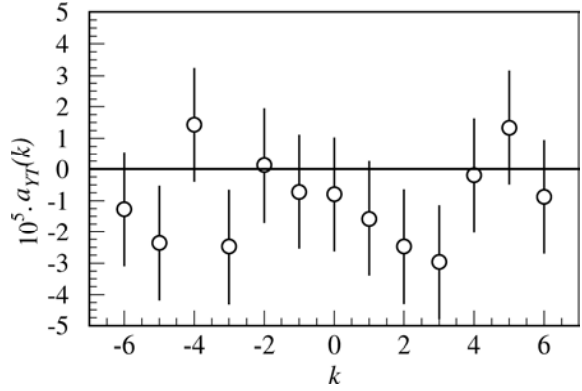

Fig. 8. Cross-correlation coefficients of spatial and temporal bit strings in the COMBO quantum random number generator.

Finally, we obtain $\hat{b}_C = (7.1 \pm 1.6) \cdot 10^{-8}$, $\hat{a}_C = (3.7 \pm 1.2) \cdot 10^{-12}$. In order to measure errors of that magnitude with 95% C.L. one would need to generate a string of at least $6 \cdot 10^{13}$ bits. Thus we prove that any generated string shorter than that is indistinguishable from a true random string and in that case there is no need to do any further tests of the generated output. Nevertheless, as a cross-check, we perform "standard" statistical analysis of sequences of $10^9$ bits by means of the NIST statistical test suite (STS, version 2.2.1) [20] using the default test parameters. Results obtained for a typical sequence are shown in Table 1.

**Table 1. Typical results of the NIST Statistical Test Suite**

| Statistical test | p-value | Proportion/Threshold | Pass |
|---|---|---|---|
| Frequency | 0.75186 | 991/980 | Yes |
| Block frequency | 0.82372 | 991/980 | Yes |
| Cumulative sums | 0.63859 | 992/980 | Yes |
| Runs | 0.67868 | 991/980 | Yes |
| LongestRun | 0.48464 | 982/980 | Yes |
| Rank | 0.35864 | 994/980 | Yes |
| FFT | 0.95120 | 992/980 | Yes |
| NonOverlappingTemplate | 0.44015 | 990/980 | Yes |
| OverlappingTemplate | 0.18454 | 992/980 | Yes |
| Universal | 0.97305 | 989/980 | Yes |
| ApproximateEntropy | 0.56463 | 992/980 | Yes |
| RandomExcursions | 0.48645 | 628/622 | Yes |
| RandomExcursionsVariant | 0.44883 | 628/622 | Yes |
| Serial | 0.47577 | 989/980 | Yes |
| LinearComplexity | 0.72582 | 989/980 | Yes |

## 5. Resilience to hardware failure and signal injection attack

When a RNG is used in a critical application (e.g. for cryptographic security) it is important that it possess a resilience to common hardware failure and allows for robust monitoring of its proper functioning. For COMBO RNG, in case that one of the detectors (D0, D1) fails completely (i.e. stops generating pulses), the BSR circuit (Fig. 2) will be either stuck to logic 1 or logic 0, depending on whether D0 or D1 failed, respectively. However the T1T2 section will still function normally albeit at only a half rate. As a result, randomness of the output will be slightly reduced (to that of the T1T2 section) while the bit production rate will be halved. In case of partial reduction of detection rate of one of detectors, BSR section will generate a heavily biased but still Markov sequence. In that case, according to Eqs. (8) and (9), randomness of output bits should be better that of the T1T2 method alone. This is indeed confirmed by an experiment in which average detection frequency of D0 was kept fixed at (10.00 ± 0.05) Mcps while that of D1 ($f_{D1}$) was varied from zero to 10 Mcps. Results plotted in Fig. 9. show that, due to rate mismatch between D0 and D1, the BSR section generates highly biased and correlated output which improves as $f_{D1}$ approaches $f_{D0}$. The bias $b_Y$ is out of the scope for all points except for $f_{D1} = 10$ MHz while autocorrelation $a_T$ is shown. Even so, randomness merits of the combined output, namely $a_C$ and $b_C$, are improved with respect to those of the T1T2 section ($a_T$, $b_T$) except for $f_{D1} = 0$ where the two are equal as expected. We further note that if both detectors would fail completely, the T1T2 section would not generate any Strobe signals and there would be no output. Therefore, the COMBO RNG is robust against detector(s) failure in the sense that as long as there is any output, its randomness quality is good.

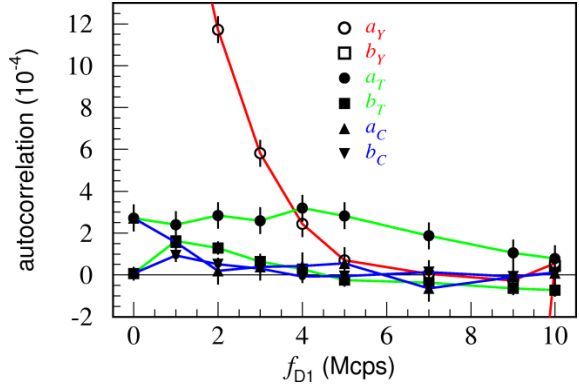

Fig. 9. Randomness merits (bias and autocorrelation) of the three sequences generated by the COMBO RNG in case when one detector is failing. Detection rate of one detector (D0) was kept at its nominal rate of 10 Mcps, while rate of the other detector (D1) was varied from 10 Mcps down to zero simulating its failure.

Let us now consider the COMBO RNG under attack by injection of a periodic signal which causes fake photon detections or "injection events" in detectors. Such an attack could be mounted by strong electromagnetic pulses or light flashes. We assume that the injected pulses affects equally and simultaneously both detectors. This would in particular be the case for a RNG miniaturized to a chip level with the closely spaced of nearly identical characteristics. We simulate the attack by mixing the CW light with strong laser pulses (PicoQuant PDL 800-D with laser head 39 ps FWHM, $\lambda = 676$ nm) via an additional beamsplitter in front of the of the RNG setup (Fig. 6). Detection probability of a laser pulse is greater than 99.7%. In an experiment the intensity of CW light was set such that $f_{D0} = f_{D1} = 10.00 \pm 0.05$ Mcps while the periodic frequency of injected laser pulses ($f_{Inject}$) was varied from zero to 7 MHz. Results are shown in Fig. 10. The attack generates a deviation from otherwise exponential time interval distribution of detected events (consisting of intertwined detected photons and injection events). Since BSR section is not sensitive to the time information it is quite insensitive to the attack. We see that the randomness merits of the output string **C** stay zero within statistical errors for $10^9$ bits up to 7 MHz. However by examining component strings **Y** and **T** we also see that randomness does deteriorate with $f_{Inject}$ albeit this would be that is apparent at higher yet statistics. Namely, every injected signal generates one additional "1" in the string **S**. This effect becomes apparent for $f_{Inject} > 4.5$ MHz when the bias of the string **Y** becomes negative while autocorrelation becomes non-Markov. The T1T2 section is by definition more sensitive to time interval distribution and that is clearly visible through increasingly negative bias and sharp rise of autocorrelation for $f_{Inject} > 3$ MHz.

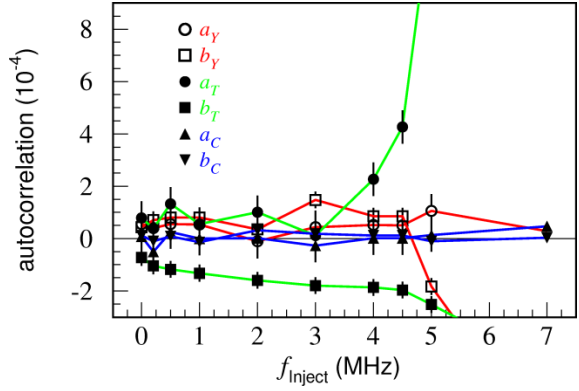

Fig. 10. Randomness merits (bias and autocorrelation) of the three sequences **Y**, **T** and **C** generated by the COMBO RNG under attack by injection of a periodic signal which causes simultaneous fake photon detections in detector with a frequency $f_{\text{Inject}}$.

In general, favorable interplay of the two sections (BSR and T1T2) makes the COMBO RNG surprisingly robust against detector failure and signal injection attacks. However, some deterioration does appear in case of strong deterioration or strong attack conditions. For practical applications it would therefore be important if failure and attacks could be detected at levels that are far from causing any noticeable randomness deterioration. We studied two simple measures: bit generation rate ($f_G$) and blanked events rate ($f_B$). Blanked events are obtained by logic AND-ing input and output of the blanking circuit in the BSR circuit shown in Fig. 2. These events are *not* used for random number generation. In Fig. 11 $f_G$ (shown by quadratic markers) and $f_B$ (shown by round markers) are plotted for three scenarios studied above: (*i*) D0 and D1 failing simultaneously; (*ii*) only D0 failing; and (*iii*) signal injection attack. For the first scenario (dotted line) $f_D$ is detection rate of both detectors; for the second (dashed line) $f_D$ is the detection rate of detector D0; and for the third (full line) $f_D$ is the frequency of injected laser pulses.

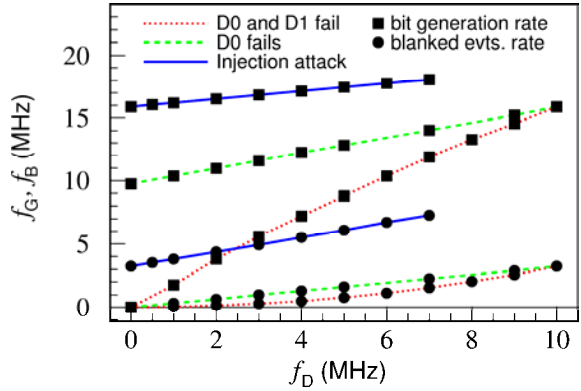

Fig. 11. Bit generation rate (quadratic dots) and blanked events rate (round dots) as functions of detection rates in three failure and attack scenarios (see the text). Both rates are sensitive to irregular operation of the RNG this allowing for robust monitoring and failure/attack detection.

We see that both generation ($f_G$) and blanked events ($f_B$) rates are quite sensitive to irregular operation of the RNG with possible advantage of $f_G$ for scenario *i* and $f_B$ for scenario *iii* whereas either is an equally good measure for scenario *ii*. In the failure scenarios (*i* and *ii*) both parameters ($f_G, f_B$) drop below normal values whereas in the attack scenario (*iii*) both measures rise above their normal values. We thus have two measures, either of which (but preferably both) can be used to robustly alert malfunction of the COMBO RNG at levels at which randomness is not noticeably compromised.

## 6. Conclusion

A RNG based on quantum effects in photonic emission and detection is presented. A mathematical framework for estimation of randomness quality based on simple measurements has been developed. The RNG is unique in several aspects: (1) the method of extraction of random bits simultaneously uses both spatial and temporal quantum information contained in the system; (2) the RNG is robust against hardware failure and signal injection attack in the sense that up to some threshold levels of detector failure or attack frequency randomness is virtually intact; (3) malfunction of the RNG, due to detector failure or signal injection, can be robustly detected at levels at which randomness is not significantly affected thus enabling protection of integrity of generated random bits. It is also shown that generated numbers pass the NIST Statistical Test Suite (STS) without post-processing. Having in mind that partial detector failure scenario is identical to an initial or aging-related difference between detectors, one can conclude that COMBO RBG is also robust on initial components variation and aging, which makes it a good candidate for mass-production or chip-level QRNG technology.

## 9. Acknowledgements

This work was supported by Fulbright program (academic year 2010/2011),, Ministry of Science Education and Sports of Republic of Croatia (contract Nr. 098-0352851-2873). Extensive numerical simulations and randomness tests were made at Center for Scientific Computing UC Santa Barbara (contract Nr. NSF-CNS-0960316) and University Computing Centre (SRCE) of University of Zagreb.